\documentclass[aps,a4paper,superscriptaddress,nofootinbib,12pt]{revtex4}
\usepackage[T1]{fontenc}
\usepackage{hyperref}
\usepackage{amsmath}
\usepackage{amsthm}
\usepackage{amsfonts}
\usepackage{xcolor}
\usepackage{amssymb}
\usepackage{stmaryrd}
\usepackage[toc,page]{appendix}
\newcommand{\comment}[1]
{{\bfseries \color{red} #1}}

\numberwithin{equation}{section} 

\DeclareMathOperator{\arccosh}{arcCosh}
\DeclareMathOperator{\arcsinh}{arcSinh}

\def\be{\begin{equation}}
\def\ee{\end{equation}}
\def\bea{\begin{eqnarray}}
\def\eea{\end{eqnarray}}

\begin{document}

\def\ra{\longrightarrow}
\def\ci{{\cal I}}
\def\ca{{\cal A}}
\def\cb{{\cal B}}
\def\cc{{\cal C}}
\def\cd{{\cal D}}
\def\cm{{\cal M}}
\def\cv{{\cal V}}
\def\cw{{\cal W}}
\def\cR{{\cal R}}
\def\hr{{\hat R}}

\title{New class of hybrid metric-Palatini scalar-tensor theories of gravity}

%
 
\author{A. Borowiec and A. Kozak}
	\email{andrzej.borowiec@uwr.edu.pl, aleksander.kozak@uwr.edu.pl}
\affiliation{$^1$ Institute of Theoretical Physics University of Wroclaw, pl. Maxa Borna 9, 50-206 Wroclaw, Poland\\
}


\begin{abstract}

A class of scalar-tensor theories (STT) including a non-metricity that unifies metric, Palatini and hybrid metric-Palatini gravitational actions  with non-minimal interaction is proposed and investigated from the point of view of their consistency with generalized conformal transformations. It is shown that every such theory can be represented on-shell by a purely metric STT possessing the same solutions for a  metric and a scalar field. A set of generalized invariants is also proposed. This extends the formalism previously introduced in \cite{kozak2019}. We then apply the formalism to Starobinsky model, write down the Friedmann equations for three possible cases: metric, Palatini and hybrid metric-Palatini, and compare some inflationary observables.
\end{abstract}


\maketitle



\bibliographystyle{plain}
\maketitle
\section{Introduction}
$F(R)$ theories of gravity have been conceived as the simplest modification of Einstein's General Relativity (GR) \cite{capo2011}-\cite{so2}. The modification is achieved by a straightforward replacement of the Einstein-Hilbert action with a function of the curvature scalar, and the main aim of such an alternation is to create a theory which would encompass phenomena that cannot be satisfactorily explained by GR only, such as the accelerated expansion of the universe \cite{carroll2004} - \cite{sergei2017}. The present-day cosmic speed-up is explained by the presence of a cosmological constant $\Lambda$, accounting for the 'dark' energy content, amounting to as much as 68.3\% of the total matter-energy density \cite{clifton2015}. The exact nature of the dark energy remains still unknown. The cosmic speed-up might be also explained by a modification of General Relativity different from adding the cosmological constant. It is possible to obtain such a behavior of the universe by including corrections in the Einstein-Hilbert action. $F(R)$ theories proved useful also in the context of cosmic inflation, i.e. an epoch that occurred shortly after Big Bang, during which the universe underwent an accelerated expansion, which led to the observed homogeneity and resolved the flatness problem \cite{staro1980}-\cite{tt4} . An $F(R)$ model remaining with a very good agreement with the Planck satellite observations is the Starobinsky model, in which the Einstein-Hilbert action was supplemented with a quadratic correction \cite{staro1980}.

So far, $F(R)$ theories have been mostly analyzed in the metric \cite{sot2006}-\cite{cap2008} and Palatini approaches \cite{bor2006}-\cite{allemandi2006}. In the metric approach, one treats the metric field as the only dynamical variable entering the action, whereas in the Palatini approach the connection is now independent of the metric tensor, and the field equations are obtained by performing variation with respect to both the metric and the connection. Based on the equations, one determines the relation between these two objects. In the case of $F(R)$ theories, the connection turns out to be an auxiliary field, and the theory becomes effectively metric. An interesting feature of $F(R)$ gravity is its equivalence to some classes of scalar-tensor theory. By performing a Legendre transformation, one can introduce a scalar field non-minimally coupled to the curvature, and analyze the theory using mathematical machinery developed for scalar-tensor gravity. It turns out that in case of Palatini $F(R)$ theory, unlike in the metric version, the scalar field has no dynamics, which means that the formalism introduces no additional degree of freedom.

An interesting generalization of the Palatini and metric $F(R)$ theories of gravity are so-called hybrid metric-Palatini theories, which were devised to avoid certain shortcomings manifested by both theories \cite{capozz2015}-\cite{harko2012}. For example, metric $F(R)$ theories introduce an additional degree of freedom behaving as a scalar field. To have an impact on large scales, it should
have a low mass. Presence of such field, however, affects dynamics at a shorter scale as well, and it should be possible to observe its infuence on our Solar System. Because no such effect was observed, one must introduce a screening mechanism \cite{capozziello2008b}, \cite{khoury2004}. On the other hand, the field is just an algebraic function of the trace of the energy-momentum tensor, so that it introduces no additional degrees of freedom \cite{olmo2011}. This leads to very serious drawbacks, for example to infinite tidal forces at surfaces of compact objects. The hybrid metric-Palatini theory, however, introduces long-range forces without being in conflict with local measurements and the need to invoke screening mechanisms. It also predicts viable formation of large-scale structures in accelerating cosmologies \cite{capozz2015}, \cite{capozziello2013}, \cite{lima2014}. 

All issues described above are usually discussed in a more general framework 
of STT (see also \cite{lind1976}-\cite{koivisto2006}) to which any $F(R)$ theory can be transform. Metric, Palatini and hybrid metric-Palatini theories have all a scalar-tensor representation, which, for the hybrid case, will be presented in the first section of the paper. This is not, however, equivalent to saying that any scalar-tensor theory arises from some $F(R)$ gravity. For such an equivalence (in a mathematical sense) to be present, certain conditions must be satisfied. 

Our idea is to present a new approach to scalar-tensor theories of gravity that unifies three previously investigated in the literature: metric, Palatini, and hybrid. Such an approach will encompass within one family of theories not only metric, but also Palatini scalar-tensor theories of gravity, and will be a natural extension of the hybrid metric-Palatini gravity. The proposed formalism will also allow one to determine if a given STT is equivalent to some metric, Palatini or hybrid $F(R)$ gravity.

The paper is organized as follows: in the first section, we discuss shortly how one can obtain scalar-tensor representation of $F(R)$ gravity. Next, we review hybrid metric-Palatini theories and then present their generalization, postulating an action functional, writing the equations of motion and solving for connection. In the last section, we switch our attention to cosmological applications of the theory and write Friedmann equations for metric, Palatini and hybrid $F(R)$ theories. As an example, we analyze the Starobinsky model and compare inflationary parameters. Some more technical aspects are presented in two Appendices. The first one is focused on formal properties of the Legendre transformation and some (partially new) examples of inflationary potentials. The second one extends the formalism of frame transformations and their invariants \cite{kozak2019} to a new hybrid metric-Palatini STT case.

Throughout the paper we work with general spacetime of dimension $n$ with a metric of the Lorentzian signature $(-,+,\ldots,+)$. We stick to the following convention when writing the curvature scalar: $R$ is a general curvature, $\mathcal{R}$ denotes curvature built from the metric only, and $\hat{R}$ is Palatini curvature, i.e. constructed both from the metric and connection. Our notational conventions are borrowed from \cite{kozak2019}.
 
 \section{From $F(R)$ to scalar-tensor gravity}

In this subsection we review  known facts concerning 
  metric, Palatini as well as hybrid $F(R)$-gravity ( see e.g. \cite{capozz2015}-\cite{harko2012}).

Consider the action of minimally coupled $F(R)$-gravity:
\begin{equation}\label{Paction}
S_F[g_{\mu\nu}, .]=\frac{1}{2\kappa^2}\int_\Omega\mathrm{d}^nx\sqrt{-g}F( R)+ S_\text{matter}(g_{\mu\nu},\chi),
\end{equation}
where $F(R)$ is a function of either a  Ricci $R=\cR(g)$ or a Palatini-Ricci $R=\hr\equiv g^{\mu\nu}\hr_{\mu\nu}(\Gamma)$ scalar and $\Gamma$ denotes torsionless connection. The matter part of the action $S_\text{matter}$ is assumed to be metric-dependent (independent of the connection).	

In both cases the action (\ref{Paction}) is dynamically equivalent to the constrained system with linear gravitational Lagrangian \footnote{One should stress that Palatini $F(R)$-gravity is not dynamically equivalent to a metric one with the same function $F(R)$.}

\begin{equation}\label{action1}
S[g_{\mu\nu}, . , \Xi]=\frac{1}{2\kappa^2}\int_\Omega\mathrm{d}^nx\sqrt{-g}\left(F^\prime(\Xi)( R-\Xi) + F(\Xi) \right) + S_\text{matter}(g_{\mu\nu},\chi).
\end{equation}

Introducing further a scalar field $\Phi=F'(\Xi)$ and taking into account the constraint equation $\Xi=R$, one arrives at the dynamically equivalent STT action  with a non-dynamical scalar field
\begin{equation}\label{actionF}
S[g_{\mu\nu}, . ,\Phi]=\frac{1}{2\kappa^2}\int_\Omega\mathrm{d}^nx\sqrt{-g}\left(\Phi R - U_F(\Phi) \right)+ S_\text{matter}(g_{\mu\nu},\chi) 
\end{equation}
either in metric or Palatini case. The potential:  
\begin{equation}\label{PotentialP}
U_F(\Phi)\equiv R(\Phi)\Phi-F(R(\Phi))\,,
\end{equation}
is the result of Legendre transformation (see Appendix A for details),
where $\Phi = \frac{d F(R)}{d R}$ and $ R\equiv \Xi = \frac{d U_F(\Phi)}{d\Phi}$.

It is known that both cases can be realized as Brans-Dicke (BD) theories with different values of the parameter $\omega_{BD}$. Original  BD is a metric scalar-tensor theory determined by the gravitational action:
\begin{equation}\label{actionBD}
S_{BD}[g_{\mu\nu}, \Phi] =\frac{1}{2\kappa^2}\int_\Omega\mathrm{d}^nx\sqrt{-g}\left(\Phi R -\frac{\omega_{BD}}{\Phi}\partial_\mu\Phi\partial^\mu\Phi- U(\Phi) \right),
\end{equation}		
where BD parameter  $\omega_{BD}\in \mathbb{R}$ and $U(\Phi)$ denotes the self-interaction potential.		
 
The action (\ref{actionBD}) is cast in so-called Jordan frame. The Jordan frame is characterized by a non-minimal coupling between the curvature and the scalar field, with the matter part of the action depending on the metric and matter fields only. One may also make use of a conformal transformation of the metric tensor, defined by (\ref{e1}), in order to switch to a frame in which the theory will be easier to analyze. Most commonly, one will choose the transforming function $\gamma_1$ in such a way, that the curvature and scalar field will no longer be coupled. One calls such frame the 'Einstein frame'. This, however, comes at a price of an anomalous coupling between the scalar and matter fields, leading to a violation of the Weak Equivalence Principle. The issue of whether Jordan or Einstein frame is the physical one remains open and has been widely discussed in the literature \cite{kuusk2015}-\cite{burns2016}.

\section{Hybrid metric-Palatini theory}

In the hybrid metric-Palatini theory, one adds to the metric Einstein-Hilbert action a function of the Palatini curvature scalar $\hat{R}(g, \Gamma)$, which can be treated as a correction term \cite{capozz2015}. The theory was devised to serve as a bridge between metric and Palatini theories, allowing one to avoid certain drawbacks of the latter. 

The action functional is given by \cite{harko2012}:
\begin{equation}\label{hybridAction}
S[g_{\mu\nu}, \Gamma^\alpha_{\mu\nu}] = \frac{1}{2\kappa^2}\int_\Omega d^nx \sqrt{-g} [\Omega_A \mathcal{R}(g) + F(\hat{R}(g, \Gamma))] + S_\text{matter}[g_{\mu\nu}, \chi]
\end{equation}

where $\Omega_A$ is a coupling constant.

We will be interested in the scalar-tensor representation of the theory. In order to switch to desired form of the action, we follow the standard procedure and perform a Legendre transformation of the $F(\hat{R})$ function, introducing a scalar field (see Eq. (\ref{action1})) and defining a potential $U_F(\Phi)$ as in \eqref{PotentialP} (cf. \ref{pot}). We will arrive at the following form of the action functional:

\begin{equation}\label{hybridAction2}
S[g_{\mu\nu}, \Gamma^\alpha_{\mu\nu}, \Phi] = \frac{1}{2\kappa^2}\int_\Omega d^nx \sqrt{-g} [\Omega_A \mathcal{R}(g) + \Phi \hat{R}(g, \Gamma) - U_F(\Phi)] + S_\text{matter}[g_{\mu\nu}, \chi]
\end{equation}

It is clear now that variation w.r.t. the connection will produce exactly the same result as in case of purely Palatini $F(\hat{R})$ gravity, i.e. the curvature scalar $\hat{R}$ will turn out to be a function of the conformally related metric $\bar{g}_{\mu\nu} = \Phi^{\frac{2}{n-2}} g_{\mu\nu}$. Therefore, we can use the result of \cite{capozz2015} and express the action (\ref{hybridAction2}
) as a function of the metric and scalar field only \footnote{The generalization of this transformation will be introduced in the next Section.}:
\begin{equation}\label{hybridMetric}
\begin{split}
S[g_{\mu\nu}, \Phi]&=\frac{1}{2\kappa^2}\int_\Omega\mathrm{d}^nx\sqrt{-g}\left((\Omega_A + \Phi) \mathcal{R}(g) +\frac{n-1}{(n-2)\Phi}\partial_\mu\Phi\partial^\mu\Phi - U_F(\Phi) \right)\\
&+S_\text{matter}[g_{\mu\nu},\chi].
\end{split}
\end{equation}
Let us now perform a shift (re-definition, cf. \eqref{e3}) in the scalar field and introduce $\psi = \Omega_A + \Phi$; this will yield:
\begin{equation}\label{hybridBD}
\begin{split}
S[g_{\mu\nu}, \psi]&=\frac{1}{2\kappa^2}\int_\Omega\mathrm{d}^nx\sqrt{-g}\left(\psi \mathcal{R}(g) - \frac{\omega_{BD}(\psi)}{\psi}\partial_\mu\psi\partial^\mu\psi - U_F(\psi) \right)\\
&+S_\text{matter}[g_{\mu\nu},\chi].
\end{split}
\end{equation}
where the Brans-Dicke parameter $\omega_{BD}$ is now a function of the scalar field:
$$\omega_{BD}(\psi) = - \frac{(n-1) \psi}{(n-2)( \psi - \Omega_A)}$$

One can observe that when $\Omega_A \rightarrow 0$ then the theory becomes Palatini $F(\hat{R})$ gravity. In the limit $\Omega_A \rightarrow \infty$, however, it reproduces GR. Any value of the parameter $\Omega_A$ lying in between these two values gives a mixture of the two approaches \cite{capozz2015}.

Let us now compute the integral invariant $\mathcal{I}^n_M$  as defined in \cite{kuusk2015} (cf. its generalization presented in  Appendix \ref{AppendixB}):
\footnote{Such invariants are determined up to an integration constant and can be normalized in various ways.}
\begin{equation}\label{I3} 
\mathcal{I}^n_M(\psi)= \frac{1}{\sqrt{n}} \int^\psi_{\psi_0}\sqrt{\pm\frac{(n-2)\mathcal{A}(\psi')\mathcal{B}(\psi')+(n-1)(\mathcal{A}'(\psi'))^2}{\mathcal{A}^2(\psi')}}d\psi' 
\end{equation}
for this theory, characterized by the following set of functions of the scalar field: $(\mathcal{A}(\psi) = \psi, \mathcal{B}(\psi) = -\frac{(n-1)}{(n-2)(-\Omega_A + \psi)}, \mathcal{V}(\psi) = V(\psi), \alpha(\psi) = 0)$. The invarant is given by:
$$\mathcal{I}^n_M(\psi) = \sqrt{\frac{4(n-1)}{n}}\Bigg[\text{arctan}\left(\sqrt{\frac{\Omega_A - \psi}{\Omega_A}}\right) - \text{arctan}\left(\sqrt{\frac{\Omega_A - \psi_0}{\Omega_A}}\right)\Bigg]\,.$$
For the metric $F(\mathcal{R})$ theory, the invariant takes the following form:
\begin{equation}
\mathcal{I}^n_M(\psi) = \sqrt{\frac{n-1}{n}}\ln \left(\frac{\psi}{\psi_0}\right),
\end{equation}
and for the Palatini $F(\hat{R})$:
\begin{equation}
\mathcal{I}^n_M(\psi) = 0\,.
\end{equation}
Three different values of this invariant enable us to distinguish between three different cases of $F(R)$-gravity (hybrid, Palatini and metric) in a frame independent way. This is due to the fact that transformations \eqref{e1}, \eqref{e3} do not change invariant quantities characterizing metric STT's.

\section{Hybrid metric-Palatini generalization}
In order to encompass both metric and Palatini scalar-tensor theories of gravity within one hybrid approach, we postulate the following action functional: 
\begin{equation}
\begin{split}
S[g_{\mu\nu},\Gamma^\alpha_{\mu\nu},\Phi]&=\frac{1}{2\kappa^2}\int_{\Omega}d^nx\sqrt{-g}\Big[\mathcal{A}_1(\Phi)\cR(g)+\mathcal{A}_2(\Phi)\hr(g,\Gamma)-\mathcal{B}(\Phi)g^{\mu\nu}\partial_\mu\Phi\partial_\nu\Phi \\
&-Q^\mu(g,\Gamma)\mathcal{C}_1(\Phi)\partial_\mu\Phi
-\bar{Q}^\mu(g,\Gamma)\mathcal{C}_2(\Phi)\partial_\mu\Phi-\mathcal{V}(\Phi)\Big]+S_{\text{matter}}[e^{2\alpha(\Phi)}g_{\mu\nu},\chi].
\end{split} \label{actionHybrid}
\end{equation}
depending on the choice of seven functions of one-variable $\ca_1, \ca_2, \cb, \cv, \cc_1, \cc_2, \alpha$, which determine the so-called frame. 
 We stick to the convention that all quantities with 'hat' are calculated using the independent connection $\Gamma$. In particular, the quantities  $Q_\mu = g^{\alpha\beta}\hat\nabla_\mu g_{\alpha\beta}$ and $\bar{Q}_\mu = - g^{\alpha\beta}\hat\nabla_\alpha g_{\beta\mu}$ depend on the non-metricity of the connection $\Gamma$ and vanish only when $\Gamma^\alpha_{\mu\nu}=\Big\{\genfrac{}{}{0pt}{}{\alpha}{\mu\nu}\Big\}_g$, i.e. in the Levi-Civita case.

The action is covariant with respect to  generalized conformal transformations  as described in Appendix \eqref{AppendixB}. These transformations \eqref{ht1} -\eqref{t6} divide all frames into mathematically equivalent classes in such a way that  $(g, \Gamma, \Phi)$ - corresponding solutions of field equations (see below), transforms each other by  \eqref{e1}-\eqref{e3}.

In particular, setting $\mathcal{A}_1(\Phi)=\Omega_A, \mathcal{A}_2(\Phi)=\Phi, \mathcal{B}(\Phi)=\mathcal{C}_1(\Phi)=\mathcal{C}_2(\Phi)=0$ one gets the theory stemming from the typical hybrid metric-Palatini action. For $\mathcal{A}_1(\Phi)=0, \mathcal{A}_2(\Phi)>0$ we can recover Palatini STT class. On the other hand, setting $\mathcal{A}_1(\Phi)>0, \mathcal{A}_2(\Phi)=\mathcal{C}_1(\Phi)=\mathcal{C}_2(\Phi)=0$ we are in a purely metric subclass. Moreover, the action (\ref{actionHybrid}) is preserved under the generalized conformal transformations (\ref{e1})-(\ref{e3}) (see Appendix). 

The equations of motion are obtained by varying w.r.t. the independent variables: metric, connection and scalar field. 
The metric equations of motion have the following form: 
\begin{equation}\label{ak1}
\begin{split}
& \mathcal{A}_1(\Phi) \mathcal{G}_{\mu\nu}(g) + \mathcal{A}_2(\Phi)\hat{G}_{\mu\nu}(g,\Gamma) + \big(\mathcal{A}''_1(\Phi) + \frac{1}{2}\mathcal{B}(\Phi) - \mathcal{C}'_1(\Phi)\big)(\partial \Phi)^2 g_{\mu\nu} +
{1\over 2}\cv (\Phi) g_{\mu\nu}\\
&- \big(\mathcal{A}''_1(\Phi) + \mathcal{B}(\Phi) - \mathcal{C}'_2(\Phi)\big)\partial_\mu \Phi \partial_\nu \Phi + (\mathcal{C}_2(\Phi)\hat{\nabla}_\mu\partial_\nu - \mathcal{C}_1(\Phi)g_{\mu\nu}\hat{\Box})\Phi - \mathcal{A}'_1(\Phi)(\nabla^g_\mu\partial_\nu - g_{\mu\nu}\Box^g)\Phi \\
& +Q_{\beta\lambda\zeta} \Big[\frac{1}{2}\mathcal{C}_2(\Phi)\delta^\sigma_{(\nu}\delta^\beta_{\mu)}g^{\lambda\zeta}-\mathcal{C}_1(\Phi)\left(\frac{1}{2}g_{\mu\nu}g^{\sigma\beta}g^{\lambda\zeta}-g_{\mu\nu}g^{\sigma\lambda}g^{\beta\zeta}+\delta^\sigma_{(\mu}\delta^\beta_{\nu)}g^{\lambda\zeta}\right)\Big]\partial_\sigma\Phi  = \kappa^2 T_{\mu\nu}.
\end{split}
\end{equation}
\noindent where $T_{\mu\nu}=-\frac{2}{\sqrt{-g}}\frac{\delta S_{\text{matter}}}{\delta g^{\mu\nu}}$ denotes simply the matter stress-energy contribution and $\hat{\Box}=g^{\mu\nu}\hat\nabla_\mu \hat\nabla_\nu$. 
 
The scalar field equation of motion reads as:
\begin{equation}\label{ak2}
	\begin{split}
	& \mathcal{A}'_1(\Phi) \mathcal{R}(g) + \mathcal{A}'_2(\Phi)\hat{R}(g,\Gamma)+\mathcal{B}'(\Phi)g^{\mu\nu}\partial_\mu\Phi\partial_\nu\Phi+2\mathcal{B}(\Phi)\Box^g\Phi+2\mathcal{B}(\Phi)\partial_\mu\Phi Q_{\nu\alpha\beta}\\
	&\times\left(\frac{1}{2}g^{\mu\nu}g^{\alpha\beta}-g^{\alpha\mu}g^{\beta\nu}\right)
	 +\mathcal{C}_1(\Phi)\nabla^g_\mu Q^\mu+\mathcal{C}_2(\Phi)
	 \nabla^g_\mu\bar{Q}^\mu-\mathcal{V}'(\Phi)=-2\kappa^2\alpha'(\Phi)T,
	\end{split}
	\end{equation}
where the right hand side is due to non-minimal coupling in the action and $T=g^{\mu\nu}T_{\mu\nu}$.

The last set of equations comes from varying with respect to the connection:
\begin{equation}\label{ak3}
	\begin{split}
	& \hat{\nabla}_\alpha\left[\sqrt{-g}\left(g^{\alpha(\zeta}\delta^{\lambda)}_{\beta}-g^{\lambda\zeta}\delta^\alpha_{\beta}\right)\right]=\\
	&=\sqrt{-g}\partial_\alpha\Phi\left[g^{\alpha(\zeta}\delta^{\lambda)}_{\beta}\left(\frac{\mathcal{C}_2(\Phi)-2\mathcal{C}_1(\Phi)-\mathcal{A}_2'(\Phi)}{\mathcal{A}_2(\Phi)}\right)-g^{\lambda\zeta}\delta^\alpha_{\beta}\left(\frac{-\mathcal{C}_2(\Phi)-\mathcal{A}_2'(\Phi)}{\mathcal{A}_2(\Phi)}\right)\right].
	\end{split}
	\end{equation}
It admits the following generic solutions for  the connection:
\begin{equation}
		\begin{split}
		\Gamma^\alpha_{\mu\nu}=\Big\{\genfrac{}{}{0pt}{}{\alpha}{\mu\nu}\Big\}_g+
		2\,\mathcal{F}_1(\Phi)\delta^\alpha_{(\mu}\partial_{\nu)}\Phi-\mathcal{F}_2(\Phi)
		g_{\mu\nu}g^{\alpha\beta}\partial_\beta \Phi ,
		\end{split}\label{sol}
\end{equation}
and the non-metricity : 
$$Q_\alpha^{\mu\nu}=\nabla_\alpha g^{\mu\nu}=2\,\big(\mathcal{F}_1(\Phi)-\mathcal{F}_2(\Phi)\big)\delta_\alpha^{(\mu}g^{\nu)\rho}
\partial_\rho\Phi+2\,\mathcal{F}_1(\Phi) g^{\mu\nu}\partial_\alpha \Phi ,$$ 
where
	$$\mathcal{F}_1(\Phi)=
	\frac{2\mathcal{C}_1(\Phi)+(n-3)\mathcal{C}_2(\Phi)+(n-1)\mathcal{A}_2'(\Phi)}{\mathcal{A}_2(\Phi)(n-1)(n-2)} ,$$	
	and
$$\mathcal{F}_2(\Phi)=\frac{2\mathcal{C}_1(\Phi)-\mathcal{C}_2(\Phi)+\mathcal{A}_2'(\Phi)}{\mathcal{A}_2(\Phi)(n-2)} .$$ 
The fact that the connection can be expressed in terms of metric tensor and scalar field only means that it introduces no additional degrees of freedom in the theory. Particularly, the connection is dynamically (on-shell) metric (Levi-Civita) 
\footnote{More generally, the connection is Levi-Civita with respect to a  conformal metric $\bar g_{\mu\nu}=\Omega^2 g_{\mu\nu}$ iff $\cc_1=\cc_2$ which follows from the condition $\mathcal{F}_1=\mathcal{F}_2=(\ln\Omega)'$. In this case, the non-metricity $Q_\alpha^{\mu\nu}$ is of the Weyl type, i.e. $\nabla_\alpha g^{\mu\nu} = W_\alpha g^{\mu\nu}$.}
if and only if the following condition is satisfied ($\mathcal{F}_1(\Phi)=\mathcal{F}_2(\Phi)=0$)
$$\cc_1(\Phi) = \cc_2(\Phi)=-\ca_2'(\Phi)$$
Otherwise, one can always choose the parameters 
$$\gamma_2(\Phi)=-\mathcal{F}_1(\Phi),\quad \gamma_3(\Phi)=-\mathcal{F}_2(\Phi)$$ 
of the transformation \eqref{e2}
in such a way that the connection $\Gamma$ becomes metric for the original metric $g$. This changes the frame parameters $(\cb, \cc_1, \cc_2)$ to the new one (cf. \eqref{ht1}-\eqref{t6}) in sucha  way that $\bar\cc_1(\Phi) = \bar\cc_2(\Phi)=-\ca_2'(\Phi)$ and
\begin{equation}
\begin{split}
\bar{\mathcal{B}}(\Phi)&=\frac{(n-2)\mathcal{A}_2(\Phi) \mathcal{B}(\Phi) -(n-1)(\mathcal{A}_2'(\Phi))^2 +
	2\mathcal{A}_2'(\Phi)[
	\mathcal{C}_2(\Phi)  - n\mathcal{C}_1(\Phi)] }{ (n-2)\mathcal{A}_2(\Phi)} \\
&
+ \frac{ (n^2-5) \mathcal{C}_2(\Phi)^2  - 4 \mathcal{C}_1(\Phi)^2+2(4 +  n - n^2)\mathcal{C}_1(\Phi)\mathcal{C}_2(\Phi)\big)}{(n-2) (n-1) \mathcal{A}_2(\Phi)}.
\end{split} \label{def2}
\end{equation}
while the remaining  ones $(\ca_1, \ca_2, \cv, \alpha)$ are unchanged. Particularly, in the case $\cc_1(\Phi) = \cc_2(\Phi)=-\ca_2'(\Phi)$ one gets $\bar\cb(\Phi)=\cb(\Phi), \bar\cc_1 (\Phi)= \bar\cc_2(\Phi)=-\ca_2'(\Phi)$.

Therefore, one can get rid of the auxiliary connection from the action (\ref{actionHybrid}) and replace it
 (on-shell) with the Christoffel symbols and functions of scalar field and its first derivatives. This will effectively lead to a metric scalar-tensor theory of gravity described by the action:
\begin{equation}
\begin{split}
S[g_{\mu\nu},\Phi]&=\frac{1}{2\kappa^2}\int_{\Omega}d^nx\sqrt{-g}
\Big[\bar{\mathcal{A}}(\Phi)\cR(g) - \bar{\mathcal{B}}(\Phi)g^{\mu\nu}\partial_\mu\Phi\partial_\nu\Phi-\mathcal{V}(\Phi)\Big] \\
&+S_{\text{matter}}[e^{2\alpha(\Phi)}g_{\mu\nu},\chi].
\end{split} \label{actionMetric}
\end{equation}
where (now on-shell $Q_{\alpha\mu\nu}(g,\Gamma)=0$ and $\mathcal{R}(g) =\hat{R}(g,\Gamma)$):
\begin{equation}
\bar{\mathcal{A}}(\Phi) = \mathcal{A}_1(\Phi) + \mathcal{A}_2(\Phi)\,. \label{def3}
\end{equation}
The correspoding field equations in this frame can be recast into the form
\begin{eqnarray} 
\bar{\mathcal{A}}(\Phi) \mathcal{G}_{\mu\nu}(g) - (\nabla^g_\mu\nabla^g_\nu - g_{\mu\nu}\Box^g)\bar{\ca}(\Phi)= \bar{T}^\Phi_{\mu\nu}+\kappa^2 T_{\mu\nu}\label{ph1}\\
\bar{\mathcal{A}}'(\Phi) \mathcal{R}(g) +\bar{\mathcal{B}}'(\Phi) (\partial\Phi)^2
+2\bar{\mathcal{B}}(\Phi)\Box^g\Phi-\mathcal{V}'(\Phi)=-2\kappa^2\alpha'(\Phi)T \label{ph2}
\end{eqnarray}
where $(\partial\Phi)^2=g^{\mu\nu}\partial_\mu\Phi\partial_\nu\Phi$ and 
\begin{eqnarray}\label{ph3}
 \bar{T}^\Phi_{\mu\nu}=\bar{\mathcal{B}}(\Phi)\partial_\mu\Phi\partial_\nu\Phi
 -{1\over2}(\bar{\mathcal{B}}(\Phi)(\partial\Phi)^2+\mathcal{V}(\Phi))g_{\mu\nu}
 \\\nonumber
= -(\partial\Phi)^2\,\bar{\mathcal{B}}(\Phi)u_\mu\,u_\nu
-{1\over2}(\bar{\mathcal{B}}(\Phi)(\partial\Phi)^2+\mathcal{V}(\Phi))g_{\mu\nu}\,.
\end{eqnarray}
mimics a perfect fluid with velocity $u_\mu=\partial_\mu\Phi/\sqrt{-(\partial\Phi)^2}$ determined by the normalized gradient co-vector ($u_\mu u^\mu=-1$).
In fact, equations \eqref{ph1}, \eqref{ph2} can be obtained from \eqref{ak1}, \eqref{ak2} if one takes into account $\cc_1=\cc_2=-\ca'$ and $Q_{\alpha\mu\nu}=0$.
Moreover, the solutions for the metric and the scalar field in both frames \eqref{actionHybrid} and \eqref{actionMetric} are exactly the same while the solution for the connection is changed.

Another important fact related to the action \eqref{actionMetric} comes from the matter energy-momentum conservation. 
It follows from (20) that $\nabla^{g\mu}\bar{T}^\Phi_{\mu\nu}= -{1\over2}(\bar{\mathcal{A}}'(\Phi) \mathcal{R}(g)+2\kappa^2\alpha'(\Phi)T\,)\nabla^g_\nu\Phi$. Now, using the identity given in \cite{koivisto2006}:
$(\Box^g \nabla^g_\nu - \nabla^g_{\nu}\Box^g)\bar{\ca}(\Phi)=  \mathcal{R}_{\mu\nu}(g) \nabla^{g \mu}\bar{\mathcal{A}}(\Phi)$
we conclude that the matter stress-energy is conserved
\begin{equation}\label{ph4}
\nabla^{g\mu}T_{\mu\nu}=\alpha'(\Phi)T\,\nabla^g_\nu\Phi
\end{equation}
provided that there is  a minimal coupling with the matter. It means that the stress-energy tensor is conserved (on-shell) in all frames such that $\alpha'(\Phi)=0$ independently of the other frame parameters $(\ca_1, \ca_2, \cb, \cc_1, \cc_2, \cv)$. Otherwise, one can always change the metric $g_{\mu\nu}\rightarrow \bar{g}_{\mu\nu}=e^{-2\alpha(\Phi)}g_{\mu\nu}$ in order to obtained the matter stress-energy conservation $\nabla^{\bar{g}\mu} \bar{T}_{\mu\nu}=0$.

Finally, combining (20) with the trace of (19), one can replace (20) (cf. \cite{kuusk2015}) by 
\begin{eqnarray}
2[(n-1)(\bar{\mathcal{A}}'(\Phi))^2 +(n-2)\bar{\mathcal{A}}(\Phi)\bar{\mathcal{B}}(\Phi) ]\Box^g\Phi +\nonumber\\
\frac{d [(n-1)(\bar{\mathcal{A}}'(\Phi))^2 +(n-2)\bar{\mathcal{A}}(\Phi)\bar{\mathcal{B}}(\Phi) ]}{d\Phi} (\partial\Phi)^2+n\bar{\mathcal{A}}'(\Phi)\cv(\Phi)\\
-(n-2)\bar{\mathcal{A}}(\Phi)\mathcal{V}'(\Phi)=2\kappa^2\,T (\bar{\mathcal{A}}'(\Phi)-(n-2)\alpha'(\Phi) \bar{\mathcal{A}}(\Phi) )\,.\nonumber
\end{eqnarray}
It shows that the scalar field has no dynamics in  a metric frame if and only if 
\begin{equation}\label{ph5}
(n-1)(\bar{\mathcal{A}}'(\Phi))^2 +(n-2)\bar{\mathcal{A}}(\Phi)\bar{\mathcal{B}}(\Phi)=0 
\end{equation}
i.e. $ (2-n)\bar{\mathcal{B}}(\Phi)=(n-1)\bar{\mathcal{A}}'(\Phi) (\ln \bar{\mathcal{A}}(\Phi))' $.
It turns out that  this property is conformally-invariant and can be expressed by vanishing of $\frac{d\mathcal{I}_{M}}{ d\Phi}=0$, where 
 $\mathcal{I}_{M}(\Phi)$ denotes the integral invariant \eqref{I3}.  It will be shown later on that \eqref{ph5} uniquely characterizes BD models arising from the $F(\hat{R})$-Palatini action: $\mathcal{I}_{M}=\mathcal{I}_{}=\mbox{const}$ (cf. (\ref{I3g})).

We remark that the action  \eqref{actionMetric} does not remember the initial action  \eqref{actionHybrid} from which it has been obtained. In order to perform the inverse (off-shell) transformation one has to assume some function $\ca_2(\Phi)$ (or $\ca_1(\Phi)$, or the invariant $\mathcal{I}_{\ca}(\Phi)$)\footnote{This data determines uniquely the 
splitting \eqref{def3} as well as the functions $\bar\cc_1 = \bar\cc_2=-\ca_2'$ when $\bar{\ca}$ is given.}. Then applying all possible transformations \eqref{e2} one can recover all  generalized frames which project (on-shell) onto a given metric frame. In other words, the totality of all generalized frames \eqref{actionHybrid} indicates the fibred structure over the totality of metric frames \eqref{actionMetric}, where the projection map is given by the formulae \eqref{def2} and \eqref{def3}, keeping $(\cv, \alpha)$ unchanged.
Moreover, a point in  the fibre can be parametrized by three functions $(\ca_2, \cc_1, \cc_2)$ constrained by the equation \eqref{def2} or equivalently 
by $(\mathcal{I}_{\ca}, \gamma_2, \gamma_3)$. The subbundle of Palatini frames $\ca_2(\Phi)=\bar\ca(\Phi)$ is like a principal bundle with an abelian structure group provided by the two-parameter transformations \eqref{e2}.
On the other hand setting $\ca_2=0$ we find out singular metric frames ($\cc_1=\cc_2=0$ on-shell).

In this way different decompositions \eqref{def3} provide a family of different off-shell (solution equivalent) actions  in the form \eqref{actionHybrid}.

Having done the projection to the metric theory we are left with the possibility of using the conformal \eqref{e1} as well as diffeomorphism \eqref{e3} transformations in order to reach simpler (e.g. Einstein canonical) forms \cite{kuusk2015}. We recall that only the Palatini case satisfies the conformally-invariant condition (\ref{ph5}).

Let us now calculate the ($\mathcal{\bar{A}}, \mathcal{\bar{B}}$) functions for $R+F(R)$ theories of gravity. We will consider all three possible approaches: metric, Palatini and hybrid metric-Palatini. For these three theories, in the scalar-tensor representation, the frame functions of the scalar field are shown in the Table I. The potential $U_F(\Phi)$ is defined by means of the Lgenedre transformation as $U_F(\Phi) = \Phi R(\Phi) - F(R(\Phi))$, with $\Phi = dF(R)/dR$ (for more details see Appendix \ref{AppendixA}). These theories have different (not related by the transformations \eqref{e1}, \eqref{e3}) solutions since their invariants are different (cf. Appendix \ref{AppendixB}). In order to investigate the solutions for a metric and a scalar field it is convenient to switch to the corresponding metric ST representation: making use of the definitions (\ref{def2}) and (\ref{def3}) of the frame functions $\bar{\mathcal{A}}(\Phi)$ and $\bar{\mathcal{B}}(\Phi)$, one gets the corresponding metric ST representation as shown in the Table II.

\begin{table}
 \begin{tabular}{| c || c | c | c | c | c | c | c |} 
 \hline
   & $\mathcal{A}_1$ & $\mathcal{A}_2$ & $\mathcal{B}$ & $\mathcal{C}_1$& $\mathcal{C}_2 $& $\mathcal{V}$ & $\alpha$\\ [0.5ex] 
 \hline\hline
 \textbf{metric} & $\Phi$ & 0 & 0 & 0 & 0 & $U_F(\Phi-1)$ & 0\\ 
 \hline
 \textbf{Palatini} & 0 & $\Phi$ & 0 & 0  & 0 & $U_F(\Phi-1)$ & 0  \\
 \hline
 \textbf{hybrid} & $\Omega_A$ & $\Phi$ & 0 & 0 & 0 & $U_F(\Phi)$ & 0  \\ [1ex]
 \hline
\end{tabular}
\caption{Different frame parametrizations of $R+F(R)$ gravity.}
\end{table}
 
\begin{table}
 \begin{tabular}{| c || c | c | c | c |} 
 \hline
   & $\bar{\mathcal{A}}$ & $\bar{\mathcal{B}}$ & $\cv$ & $\alpha$\\ [0.5ex] 
 \hline\hline
 \textbf{metric} & $\Phi$ & 0 & $U_F(\Phi-1)$ & 0\\ 
 \hline
 \textbf{Palatini} & $\Phi$  & $-\frac{n-1}{n-2}\frac{1}{\Phi}$ & $U_F(\Phi-1)$ & 0\\
 \hline
 \textbf{hybrid} & $\Omega_A+ \Phi$ & $-\frac{n-1}{n-2}\frac{1}{\Phi}$ & $U_F(\Phi)$ & 0 \\ [1ex]
 \hline
\end{tabular}
\caption{
	The corresponding metric SST frames for three cases of 
$R+F(R)$ gravity.}
\end{table}

\section{Cosmological applications}

We may now attempt to write down Friedmann equations for the action (\ref{actionHybrid}). This task is pretty straightforward, as the theory turns out to be fully metric. The equations of motion will be the same as in case of metric scalar-tensor theories; the only difference is the definition of the parameters $(\bar{\mathcal{A}}, \bar{\mathcal{B}})$, since now they differ from the ones we started with. 

For four-diemnsional Friedmann-Robertson-Walker metric:
\begin{equation}\label{FRWmetric}
g_{\mu\nu} = \text{diag}\left(-1\,,  \frac{a^2(t) }{1 - kr^2}\,, a^2(t)r^2\,,a^2(t) r^2 \sin^2\theta\right),
\end{equation}
where $k$ is the spatial curvature, we get the following Friedmann equations (assuming that $\Phi = \Phi(t)$ and a barotropic $p = w\rho$ perfect fluid as a source):
\begin{subequations}
\begin{align}
& 3H^2  = \frac{\kappa^2 \rho}{\bar{\mathcal{A}}(\Phi)} - \frac{3k}{a^2} + \frac{1}{2}\frac{\bar{\mathcal{B}}(\Phi)}{\bar{\mathcal{A}}(\Phi)}\dot{\Phi}^2 - 3\frac{\bar{\mathcal{A}}'(\Phi)}{\bar{\mathcal{A}}(\Phi)}H\dot{\Phi} + \frac{1}{2}\frac{\mathcal{V}(\Phi)}{\bar{\mathcal{A}}(\Phi)}, \\
& 2\dot{H} + 3H^2  = -  \frac{w \kappa^2\rho}{\bar{\mathcal{A}}(\Phi)} - \frac{k}{a^2} -\frac{\bar{\mathcal{B}}(\Phi) + 2\bar{\mathcal{A}}''(\Phi)}{2\,\bar{\mathcal{A}}(\Phi)} \dot{\Phi}^2 - \frac{\bar{\mathcal{A}}'(\Phi)}{\bar{\mathcal{A}}(\Phi)}(2H\dot\Phi + \ddot{\Phi}) + \frac{\mathcal{V}(\Phi)}{2\,\bar{\mathcal{A}}(\Phi)} ,\\
\begin{split}
& \left(3(\bar{\mathcal{A}}'(\Phi))^2 + 2\bar{\mathcal{A}}(\Phi)\bar{\mathcal{B}}(\Phi)\right)\ddot{\Phi}  = -3\left(3(\bar{\mathcal{A}}'(\Phi))^2 + 2\bar{\mathcal{A}}(\Phi)\bar{\mathcal{B}}(\Phi)\right)H\dot{\Phi}  \\
&\quad\quad\quad\quad\quad\quad -\left((\bar{\mathcal{A}}(\Phi)\bar{\mathcal{B}}(\Phi))' + 3\bar{\mathcal{A}}'(\Phi)\bar{\mathcal{A}}''(\Phi)\right)\dot{\Phi}^2 + \left(2\mathcal{V}(\Phi)\bar{\mathcal{A}}'(\Phi) - \mathcal{V}'(\Phi)\bar{\mathcal{A}}(\Phi)\right) \\
& \quad\quad\quad\quad\quad\quad + \kappa^2\rho(1 - 3w)\left[\bar{\mathcal{A}}'(\Phi) - 2\alpha'(\Phi)\bar{\mathcal{A}}(\Phi)\right]. \label{ee3}
\end{split}
\end{align}
\end{subequations}

Here, prime denotes differentiaton w.r.t. the scalar field, dot - w.r.t. the cosmic time, and $H = \dot{a}/a$, as usual. 

Combining first two expressions one can infer deceleration/acceleration formula for the scale factor
\begin{equation}\label{ee2bis}
\frac{\ddot{a}}{a}\equiv\dot{H} + H^2  = -  \frac{ \kappa^2\rho(1+3w)}{6\,\bar{\mathcal{A}}(\Phi)} -\frac{\bar{2\mathcal{B}}(\Phi) + 3\bar{\mathcal{A}}''(\Phi)}{6\,\bar{\mathcal{A}}(\Phi)} \dot{\Phi}^2 - \frac{\bar{\mathcal{A}}'(\Phi)}{2\,\bar{\mathcal{A}}(\Phi)}(H\dot\Phi + \ddot{\Phi}) + \frac{\mathcal{V}(\Phi)}{6\,\bar{\mathcal{A}}(\Phi)} 
\end{equation}
It shows that only the last term (if positive) supports acceleration explicitly.
Otherwise some more complicated scenarios are needed in order to get the right hand side positive.

If we act with the covariant derivative on the energy-momentum tensor, we get:
\begin{equation}
\nabla_\mu T^{\mu\nu} = \alpha'(\Phi) T \partial^\nu \Phi
\end{equation}
If the anomalous coupling between the scalar field and the matter part of the action is not present, then the energy-momentum tensor is conserved. In this case, the energy density can be obtained by solving the following equation:
\begin{equation}
\label{rhoofa}
\dot{\rho} + 3H(1+w)\rho = 0,
\end{equation}
which gives:
\begin{equation}
\rho(a) = \rho_0 a^{-3(1+w)}
\end{equation}
where $\rho_0$ is the energy density at the present time. In fact, one can take into account more than one energy density source and write:
\begin{equation}
\rho_i(a) = \rho_{i,0} a^{-3(1+w_i)},
\end{equation}
with $i$ indexing different components of the total energy density, such as dust ($\rho_m, w=0$), radiation ($\rho_r, w=\frac{1}{3}$) or dark energy ($\rho_\Lambda, w=-1$). 

In the next section, we will demonstrate, on concrete examples of the Starobinsky model $F(R) = R + \beta R^2$, the differences between three approaches: metric, Palatini, and hybrid metric-Palatini. 

\subsection{Example: Starobinsky model}
Starobinsky model is a simple modification of the General Relativity. In this model, the Einstein-Hilbert action is supplemented with a quadratic correction. There are three possible approaches one can take in order to analyze the theory: treating the curvature as a function of the metric, of the metric and the connection, or assuming that only the correction term, $\beta R^2$, is constructed from both the metric and the connection:
\begin{itemize}
\item \textbf{Case 1:} $F(\mathcal{R}) = \mathcal{R}(g) + \beta\mathcal{R}(g)^2$ - metric;
\item \textbf{Case 2:} $F(\hat{R}) = \hat{R}(g,\Gamma) + \beta\hat{R}(g, \Gamma)^2$ - Palatini;
\item \textbf{Case 3:} $F(\mathcal{R}, \hat{R}) = \Omega_A\mathcal{R}(g) +\beta\hat{R}(g, \Gamma)^2$ - hybrid metric-Palatini.
\end{itemize}
Let us notice that it does not make much sense to analyze the case when the Einstein-Hilbert action (i.e. the curvature itself) is constructed \'{a} la Palatini, and the correction is metric, as the Palatini Einstein-Hilbert action always turns out to be fully metric.

The next step will be to transform the theory to the scalar-tensor representation. For the first two cases, the potential $U_F(\Phi) = \Phi R(\Phi) - F(R(\Phi))$ will be exactly the same, since the procedure does not differ for the metric and Palatini approaches. However, one will end up with different potential when considering the third case. 
All three cases generate the same quadratic potential which due the presence of linear term is shifted $U_F(\Phi)\rightarrow U_F(\Phi-\Omega_A)$ for the metric and the Palatini cases (cf. Appendix A)
Also, the coupling between the field and the curvature will not be the same as in the first two cases, because now the scalar field is defined as $\Phi = \frac{d F(\mathcal{R}, \hat{R})}{d\hat{R}}$. The differences and similarities between these three cases are shown in Table I. As one can see, metric and Palatini cases are almost identical, the difference being the value of the couplings $(\mathcal{A}_1, \mathcal{A}_2)$.

For the $\bar{\mathcal{A}}(\Phi)$ and $\bar{\mathcal{B}}(\Phi)$ functions shown in the Table 1., the Friedmann equations will read as follows:

\textbf{Metric:}
\begin{subequations}
\begin{align}
3H^2 & =  \frac{\kappa^2}{\Phi}\sum_i\rho_i - 3\frac{k}{a^2} -3H\frac{\dot{\Phi}}{\Phi} + \frac{1}{8\beta}\frac{(\Phi - 1)^2}{\Phi}, \\
2\dot{H} + 3H^2 & = -\frac{\kappa^2 }{\Phi}\sum_i w_i \rho_i -\frac{k}{a^2} - 2H\frac{\dot{\Phi}}{\Phi} -\frac{\ddot{\Phi}}{\Phi} + \frac{1}{8\beta}\frac{(\Phi - 1)^2}{\Phi},\\
\ddot{\Phi} & = \frac{\kappa^2}{3}\sum_i(1-3w_i)\rho_i-3H\dot{\Phi} - \frac{\Phi-1}{6\beta}.
\end{align}
\end{subequations}

\textbf{Palatini:}

In case of the Palatini approach, scalar field has no dynamics, so it introduces no additional degrees of freedom. This is caused by the fact that the denominator in the Eq. (\ref{ee3}) vanishes. The equations are given by:
\begin{subequations}
\begin{align}
3H^2 & = \frac{\kappa^2}{\Phi}\sum_i\rho_i  -3\frac{k}{a^2}-\frac{3}{4}\frac{\dot{\Phi}^2}{\Phi^2} -3H\frac{\dot{\Phi}}{\Phi} + \frac{1}{8\beta}\frac{(\Phi - 1)^2}{\Phi}, \\
2\dot{H} + 3H^2 & = -\frac{\kappa^2 }{\Phi}\sum_i w_i \rho_i  -\frac{k}{a^2}+\frac{3}{4}\frac{\dot{\Phi}^2}{\Phi^2} - 2H\frac{\dot{\Phi}}{\Phi} -\frac{\ddot{\Phi}}{\Phi} + \frac{1}{8\beta}\frac{(\Phi - 1)^2}{\Phi},\\
0 & =  \kappa^2 \sum_i(1-3w_i)\rho_i - \frac{\Phi-1}{2\beta}.
\end{align}
\end{subequations}
The third equation means that, in principle, one can express $\Phi$ in terms of $\rho_i$, as the relation between these objects is algebraic. Therefore, the evolution equation for the scale factor turns out to be of second order and can be described as a  two-dimensional dynamical system of Newtonian type with an effective potential function (cf. \cite{borowiec2012,borowiec2015,stach2017}). This becomes even more transparent in another conformally-equivalent Einstein ($\ca=1, \cb=0, \alpha ={1\over 2} \ln\Phi$) frame, cf. eq.s (27) - (28) in \cite{stach2017}.

\textbf{Hybrid metric-Palatini}
\begin{subequations}
\begin{align}
3H^2 & =  \frac{\kappa^2}{\Omega_A+\Phi}\sum_i\rho_i - 3\frac{k}{a^2}-\frac{3}{4}\frac{\dot{\Phi}^2}{\Phi(\Omega_A+\Phi)} -3H\frac{\dot{\Phi}}{\Omega_A+\Phi} + \frac{1}{8\beta}\frac{\Phi^2}{\Omega_A+\Phi}, \\
2\dot{H} + 3H^2 & = -\frac{\kappa^2 }{\Omega_A+\Phi}\sum_i w_i \rho_i -\frac{k}{a^2}+\frac{3}{4}\frac{\dot{\Phi}^2}{\Phi(\Omega_A+\Phi)} - 2H\frac{\dot{\Phi}}{\Omega_A+\Phi} -\frac{\ddot{\Phi}}{\Omega_A+\Phi} + \frac{1}{8\beta}\frac{\Phi^2}{\Omega_A+\Phi},\\
\ddot{\Phi} & = -\frac{\Phi \kappa^2}{3\Omega_A}\sum_i(1-3w_i)\rho_i-3H\dot{\Phi} +\frac{\dot{\Phi}^2}{2\Phi} + \frac{\Phi^2}{6\beta}.
\end{align}
\end{subequations}
This shows that the dynamics of the scale factor, as well as the scalar field, is different in all three cases.  Only in the Palatini case, the scalar field has no dynamics. 
More precisely, three cases are mathematically different and cannot be related by a conformal transformation of the metric and a scalar field redefinition. 
We are going to illustrate now how some physical prediction can differ for the models presented above. To this aim let us consider inflationary parameters. 

Within the context of scalar-tensor theories of gravity, where non-minimal coupling between curvature and scalar field might be present, one must keep in mind that certain physical predictions, such as the number of e-folds, may strongly depend on the choice of conformal frame. However, certain observables can be expressed in a way that is frame-independent. For example, as it was shown in \cite{karam2017}, it is possible to express slow-roll parameters characterizing inflation in a manifestly frame-independent way, making use of an invariant generalization of the scalar field potential (cf. Appendix B). There is, however, a caveat in this way of thinking. As Karam \textit{et al.} are showing in the paper \cite{karam2017}, even though the spectral indices become functions of the invariant potential, which has the same form in every conformal frame, their numerical values might be different due to the fact that in different frames, the inflation lasts for different number of e-folds. Since we are not interested here in comparing conformal frames, but rather in comparing values of spectral indices for different theories, we decide to carry out all calculations in what is called 'Einstein frame'. 

In the Palatini case, there is no additional degree of freedom related to the scalar field (as it can be expressed as a function of matter, which is negligible during the inflation), so it cannot give rise to any dynamical fluctuations. Only metric and hybrid theories will be of any interest to us 
\footnote{The method proposed below does not apply in the Palatini case where $\mathcal{I}=0$. As it was shown in our earlier papers \cite{borowiec2016,stach2017} the inflationary effects manifest themselves differently in both Jordan as well as in Einstein frame. They are the results of some singularities in an effective Newtonian-type potential  governing the universe evolution.}. We start by computing the spectral indices 
making use of notation introduced in \cite{kuusk2016}. First, one must compute the invariant potential and express it in terms of an invariant generalization of the scalar field (\ref{I3g}). Having obtained the potential, one can compute the invariant slow-roll parameters and the number of e-folds, and then substitute the result in the formula for spectral indices.

One can write the following invariant slow-roll parameters characterizing cosmic inflation \cite{kuusk2016}:
\begin{subequations}
\begin{align}
& \hat{\kappa}^{(V)}_0 = \frac{1}{4\ci_2^2}\left(\frac{d\ci_2}{d\ci}\right)^2, \\
 & \hat{\kappa}^{(V)}_1 = 4\hat{\kappa}^{(V)}_0 - \frac{1}{\ci_2}\frac{d^2\ci_2}{d\ci^2}.
\end{align}
\label{invsrp}
\end{subequations}
where in four dimensions (cf. \eqref{I3}) 
$$\mathcal{I}\equiv\mathcal{I}_M^4\,,\qquad\quad \mathcal{I}_2 = \frac{\mathcal{V}(\Phi)}{\mathcal{A}(\Phi)^2}\,.$$  
Slow-roll conditions are given by $ |\hat{\kappa}^{(V)}_i| \ll 1$, and the inflation ends when $\hat{\kappa}^{(V)}_0 = 1$.
In the beginning, we analyse the metric Starobinsky model. Invariant $\mathcal{I}$ is given by:
\begin{equation}
\mathcal{I} = \frac{\sqrt{3}}{2}\ln\left(\frac{\Phi}{\Phi_0}\right)
\end{equation}
so that
\begin{equation}
\Phi(\mathcal{I}) = \Phi_0 e^{ \frac{2}{\sqrt{3}} \mathcal{I}}
\end{equation}
The invariant potential can be written as:
\begin{equation}
\mathcal{I}_2 = \frac{\left(e^{\frac{2}{\sqrt{3}} \mathcal{I}}-1\right)^2}{4\beta e^{\frac{4}{\sqrt{3}} \mathcal{I}}}
\end{equation}
Upon substitution in (\ref{invsrp}), one gets:
\begin{equation}
\hat{\kappa}^{(V)}_0 = \frac{4}{3\left(e^{\frac{2}{\sqrt{3}} \mathcal{I}}-1\right)^2}, \quad\hat{\kappa}^{(V)}_1 = \frac{8e^{\frac{2}{\sqrt{3}} \mathcal{I}}}{3\left(e^{\frac{2}{\sqrt{3}} \mathcal{I}}-1\right)^2}
\end{equation}
The number of e-folds in the Einstein frame can be computed from the following formula:
\begin{equation}
\hat{N} = - \int_{\mathcal{I}^0}^{\mathcal{I}^\text{end}}\frac{1}{\sqrt{2\hat{\kappa}^{(V)}_0}}d\mathcal{I} \approx \frac{3}{4}e^{\frac{2}{\sqrt{3}} \mathcal{I}^0},
\end{equation}
where $\mathcal{I}^0$ is the value of the scalar field at the beginning of inflation, and $\mathcal{I}^\text{end}$ - at the end. We additionally assumed that $\mathcal{I}^\text{end} \ll \mathcal{I}^0$. 
Finally, we can compute the spectral index for the scalar field (up to the first order in slow-roll parameters):
\begin{equation}
\begin{split}
\hat{n}_s & = 1 - 2\hat{\kappa}^{(V)}_0 - \hat{\kappa}^{(V)}_1 = \frac{-5-14e^{\frac{2}{\sqrt{3}} \mathcal{I}^0} + 3e^{\frac{4}{\sqrt{3}} \mathcal{I}^0}}{3\left(e^{\frac{2}{\sqrt{3}} \mathcal{I}^0}-1\right)^2}\approx \frac{-5-56\frac{\hat{N}}{3} + 16\frac{\hat{N}^2}{3}}{3(4\frac{\hat{N}}{3}-1)^2} 
\end{split}
\end{equation}
For $\hat{N}=50$, one gets $\hat{n}_s \approx 0.958$, and for $\hat{N}=60$, $\hat{n}_s \approx 0.965$, in agreement with the Planck satellite result $n_s = 0.968 \pm 0.006$ \cite{planck2015}.

In the hybrid case, the invariant generalization of the scalar field is given by:
\begin{equation}
\ci =  \sqrt{3}\left(\arctan{\sqrt{\frac{\Phi}{\Omega_A}}} - \arctan{\sqrt{\frac{\Phi_0}{\Omega_A}}}\right),
\end{equation}
so that:
\begin{equation}
\Phi(\ci) = \Omega_A\tan^2\left({\frac{\ci}{\sqrt{3}} + \arctan{\sqrt{\frac{\Phi_0}{\Omega_A}}}}\right).
\end{equation}
The invariant potential can be expressed as:
\begin{equation}
\begin{split}
\ci_2 = \frac{\tan^4\left({\frac{\ci}{\sqrt{3}} + \arctan{\sqrt{\frac{\Phi_0}{\Omega_A}}}}\right)}{4\beta\Big(1 + \tan^2\left({\frac{\ci}{\sqrt{3}} + \arctan{\sqrt{\frac{\Phi_0}{\Omega_A}}}}\right)\Big)^2}.
\end{split}
\end{equation}
The slow-roll parameters can be now computed easily:
\begin{equation}
\hat{\kappa}^{(V)}_0 = \frac{4}{3}\cot^2\left(\frac{\ci}{\sqrt{3}} + \arctan{\sqrt{\frac{\Phi_0}{\Omega_A}}}\right), \quad
\hat{\kappa}^{(V)}_1 = \frac{4}{3}\csc^2\left(\frac{\ci}{\sqrt{3}} + \arctan{\sqrt{\frac{\Phi_0}{\Omega_A}}}\right).
\end{equation}
As we can see, already at this point we encounter a problem for the hybrid Starobinsky model. The second slow-roll parameter, $\hat{\kappa}^{(V)}_1$, is given by the squared cosecans function, which does not take values smaller than 1. Therefore, it is impossible to satisfy the condition $|\hat{\kappa}^{(V)}_1| \ll 1$, and further calculation reveals that, for the theory, the scalar spectral index is equal to $-\frac{1}{3}$, which is in a very strong disagreement with observations. Therefore, the hybrid Starobinsky model is disfavoured by experimental data. 
More detailed qualitative analysis and the confrontation with observational data will be presented elsewhere. 

\section{Conclusions and perspectives}

In this paper, we presented a possible generalization of hybrid metric-Palatini theories. Our idea was to add a function of a scalar field non-minimally coupled to the curvature built entirely from the metric tensor to an action functional for general Palatini scalar-tensor theories of gravity introduced in \cite{kozak2019}. In such a way, one will create a self-consistent theory being a minimal extension of the metric, Palatini and hybrid metric-Palatini gravity with the freedom of transforming the scalar field, metric tensor and affine connection independently, using the formulae (\ref{e1})-(\ref{e2}). Under this transformation, the action functional must remain form-invariant; to achieve this, one needs to transform the functions of the scalar field $(\mathcal{A}_1, \mathcal{A}_2, \mathcal{B}, \mathcal{C}_1, \mathcal{C}_2, \mathcal{V}, \alpha)$ defining, together with the dynamical variables $(g, \Gamma, \Phi)$, the conformal frame. Knowing how the functions transform, one is able to come up with certain combinations of them remaining invariant under the conformal change. Invariants can be used to check if two arbitrary scalar-tensor theories can be linked with the transformations (\ref{e1})-(\ref{e2}). If so, such theories should be considered mathematically equivalent by means of the generalized conformal transformation combined with a diffeomorphism of the scalar field. 

As it turned out, any hybrid scalar-tensor theory can be projected using $(\gamma_2, \gamma_3)$ functions to a theory which is fully metric, i.e. its coefficients satisfy the relation $\mathcal{C}_1(\Phi) = \mathcal{C}_2(\Phi) = -\mathcal{A}'_2(\Phi)$. The vectors $Q^\mu$ and $\bar{Q}^\mu$ built from non-metricity vanish on-shell and one gets a metric theory with the functions $(\bar{\mathcal{A}}, \bar{\mathcal{B}})$ given by (\ref{def2}) and (\ref{def3}). Conversely, if one is given a metric scalar-tensor theory, such as the Brans-Dicke theory in the metric approach, then it is impossible to reconstruct the hybrid theory equivalent to it without further specification of the value of the invariant $\mathcal{I}_\mathcal{A}$. In other words, hybrid theories that are not mathematically equivalent, i.e. have different invariants, can be dynamically equivalent to the same metric scalar-tensor theory.

The class of STT considered in the present paper is not a particular case of more general theory with two scalar fields and arbitrary functional dependence $F(\mathcal{R}, \hat R)$  presented in \cite{rosa1,rosa}  since it is singular in their terminology. It will be a task for our future investigation to analyze  $F(\mathcal{R}, \hat R)$  from the point of view of solution-equivalent classes and describe them in the form of invariants. Another future task would be to study more general nonminimal coupling that takes into account nonmetricity (cf. \cite{delhom}).
In this context, the idea of non-metricity driven inflation \cite{stelmach1991} should be reconsidered.

There are exactly-solvable cosmological models in metric STT for some special choices of potential functions (see e.g. \cite{faraoni2004,fermi2020} and references therein). They can be used to generate, by applying conformal transformations \eqref{e1}-\eqref{e1}, new exact solutions in Palatini and hybrid STT cases.

As it has been already mentioned, the solutions for the metric and for the scalar field in both frames \eqref{actionHybrid} and \eqref{actionMetric} are exactly the same while ones for the connection change according to \eqref{sol}. This property can be used to capture some dark energy effects related with galactic curves \cite{sporea2018,sporea2018b} or applied to stellar structure descriptions \cite{w1}-\cite{w4}.

This shows that there is a renewing interest and ongoing activity in applications of Palatini STT in astrophysics, which is enforced due to the recent developments in solving  dark matter, dark energy and cosmic inflation problems (cf. \cite{racioppi2017}-\cite{shimada2019} and \cite{coumbe}-\cite{shapo}).

Our finding has also practical meaning.  As shown, when applicable, it allows  the calculation and comparison of some   inflationary observables based on the metric STT. Secondly, in order to solve equations of motion in an arbitrary frame it might be more convenient to  find solutions for the metric and scalar field in the simpler projected metric frame or one of the conformally equivalent frames and then transform them to the initial frame getting a solution for the connection directly from \eqref{ak3}. So each metric ST model can be enriched by adding arbitrary nonmetricity, extending the Levi-Civita connection in a dynamical way. 
 For these reasons, the formalism introduced here allows for getting  better insight and deeper understanding of mutual relationships among different STT both on the operational as well the conceptual level.
 
\section*{Acknowledgments}
This work has been  supported by the Polish National Science Center (NCN), project UMO-2017/27/B/ST2/01902
and benefited from COST Action CA15117 (CANTATA), supported by COST
(European Cooperation in Science and Technology). We are grateful to Sergei Odintsov and Marek Szydlowski for discussions.


\appendix
\section{Scalar field potential from $F(R)$ modified gravity} \label{AppendixA}

\subsection {Few remarks on the Legendre transformation}

As it was already mentioned, in the purely gravitational  $F(R)$ action (\ref{Paction}) the scalar field potential  $U(\Phi)$ encoding information about the function $F(R)$ is given by
\begin{equation}\label{Pote}
	U_F(\Phi)\equiv U(\Phi)= R(\Phi)\Phi-F(R(\Phi))
\end{equation}
where  $\Phi = \frac{d F(R)}{dR}$ and $R(\Phi)$ denotes the inverse relation.  More exactly, for a given $F$ the potential $U_F$ is a (singular) solution of the Clairaut's differential equation \cite{kamke}: \footnote{
	A family of  real lines $U(\phi)=c\phi-F(c)$ evolving around the singular solution and parameterized by the integration constat $c$ consists of regular solutions.}  
\begin{align}\label{Clairaut}
	U(\Phi)=\Phi \frac{d U}{d\Phi} -F(\frac{d U}{d\Phi})\ .
\end{align}
In fact, due to the inverse function theorem, such differentiable solution exists around each point $F''(R)\neq 0$.
A remarkable property of such solution (\ref{Pote})  is that it can be always plotted on the $(\Phi,U)$-plane in the parametric form  $R_1<R<R_2$:  
\begin{align}\label{pot}
	\Phi&=F'(R) ,\\ U&=RF'(R)-F(R)
\end{align}
even if an explicit functional dependence $U(\Phi)$ remains unknown. Conversely, having done the potential $U(\Phi)$ one can plot as well on the $(R, F)$-plane the corresponding $F(R)$ function $\Phi_1<\Phi<\Phi_2$:
\begin{align}\label{potU}
R&=U'(\Phi) ,\\ F&=\Phi U'(\Phi)-U(\Phi)\,.
\end{align}

Moreover, the functional transformation $F(R)\mapsto U_F(\phi)$, a.k.a the Legendre transform,  possesses the following useful properties (which can be checked by straightforward calculations):
\begin{itemize}
	\item It is involutive, i.e. it is own inverse. It means that the function $F(R)$ is a Legendre transform of $U_F(\Phi)$.
	\item Trivial, i.e. constant,  potential $U(\Phi)$ corresponds to the linear Lagrangian $F(R)=b R+c$.
	\item More generally, for a given $F(R)$ and the corresponding Legendre transform $U_F(\Phi)$ one considers  $\tilde F(R)=aF(A R)+b R+c$, where $A, a, b, c$ are numerical constants
		\footnote{$c=-2\Lambda$ plays a role of cosmological constant.} and  $\tilde\Phi = \frac{d \tilde F(R)}{dR}$. Then 
	\begin{align}\label{tpot}
		\tilde U_{\tilde F}(\tilde\Phi)= a U_F\left(\frac{\tilde\Phi-b}{Aa}\right) -c\, .
	\end{align}
	\item Similarly, modifying linearly the potenial \eqref{tpot}
	\begin{equation}\label{linear}
		\tilde U_{\tilde F}(\tilde\Phi)\rightarrow \tilde U_{\tilde F}(\tilde\Phi)+B \tilde\Phi=a U_F\left(\frac{\tilde\Phi-b}{Aa}\right) + B \tilde\Phi -c\,.
	\end{equation}
	one finds that it results from
	\begin{equation}\label{gen}
		\tilde F(R) =aF(A (R-B))+b (R-B)+c \,.
	\end{equation}
	\item In the case of inverse function $F^{-1}$ one finds
	\begin{equation}\label{inverse}
		U_{F^{-1}}(\Phi)=-\Phi U_{F}(\Phi^{-1})\,
	\end{equation}
	or equivalently $F_{U^{-1}}(R)=-R F_{U}(R^{-1})$.
	\item Assuming  $F(R)=\int f^{-1}(R)dR$ we obtain 
	\begin{equation}\label{ab}
	U(\Phi)=\int f(\Phi) d\Phi  \end{equation}
	and vice versa.
\end{itemize}

For the purpose of this section  it is convenient to introduce the following terminology:
{\it
	We say that two functions are \textbf{weakly equivalent} $F\sim\tilde F$ if they differ by the linear transformation  (both dependent and independent variables) in the form (\ref{linear}).
	
	We say that two functions are \textbf{weakly related} by the Legendre transformation $F\leadsto \tilde U$ if  $F\mapsto U$ and $U \sim\tilde U$. 
}

\subsection{Some viable example}  
 
Our purpose now is to illustrate how the above works on some concrete examples.  \footnote{Shortcut symbols, e.g. LFI, DWI, refers to Table I on pages 14-16 in \cite{martin2014}, listing viable inflationary potentials discussed later in the paper.}\medskip\\  
{\bf Example 1.} (Power law Lagrangian) For $F(R)=R^p,\ p\neq 1,0$ we get 
$U(\Phi)= \frac{1}{q-1}\left(\frac{(q-1)\Phi}{q}\right)^q$, where $q=\frac{p}{p-1}$, \footnote{This is equivalent to $p=\frac{q}{q-1}$ or in more symmetric form to ${1\over p}+{1\over q}=1$. This expression is involutive and possesses the following asymptotic: $p\mapsto 0^\pm$ iff $q\mapsto 0^\mp$; $p\mapsto 1^\pm$ iff $q\mapsto \pm\infty$.} (see also DWI). 
The more general form  is $F(R)=a(A(R-B))^p +b(R-B)+c$ 
with the potential   $U(\Phi)= \frac{a}{q-1}\left(\frac{(q-1)(\Phi-B)}{q aA}\right)^q+ B\Phi+c$.
These cover the cases: LFI, SFI, CSI, IMI, BI, UHI, DSI from \cite{martin2014}. In particular, taking 
Starobinsky type Lagrangian $F(R)= R-2\Lambda+\gamma  R^2$ one finds 
$U(\Phi)={1\over 4\gamma }(\Phi-1)^2 +2\Lambda$ (see also DWI).\medskip\\
{\bf Example 2.} The exponential function $F(R)=e^R$ provides the logarithmic potential $U(\Phi)=\Phi(\ln\Phi-1)$. Thus $F(R)=ae^{A (R-B)}+b(R-B)+c$.  
leads to $U(\Phi)=\frac{\Phi-b}{A}(\ln(\frac{\Phi-b}{Aa})-1)+B\Phi -c$. It should be mentioned that the exponential gravity model has been already proposed in several papers (see e.g. \cite{sergei2017}).
\medskip\\
{\bf Example 3.} Conversely, exchanging $F$ and $U$ in the above example  one gets: $F(R)=a A (R-B)(\ln(A(R-B))-1)+c$ leads to
$U(\Phi)=a e^{\frac{(\Phi-b)}{Aa}}+B\Phi-c$ (cf. RCHI, ESI, PLI).\medskip\\
{\bf Example 4.} Replacing $\exp{R}$ by its inverse one gets $U_{\ln R}(\Phi)=1+\ln\Phi$ (cf. WRI).\medskip\\
{\bf Example 5.} If $F(R)=(R-1)e^R$ then $U(\Phi)=\Phi (W(\Phi)-1)+\frac{\Phi}{W(\Phi)}$, where $W$ denotes the Lambert $W$-function. This generalizes to 
$F(R)=(R-1)e^R$ and $U(\Phi)=\Phi (W(\Phi)-1)+\frac{\Phi}{W(\Phi)}=\Phi (W(\Phi)-1)+\exp W(\Phi)$.\medskip\\
\noindent {\bf Example 6.} It is possible to generalize the above case: $F(R)=e^{p W(R)}\left(W(R)+{1\over q}\right)$ gives $U(\Phi)=\left(\frac{\Phi}{p}\right)^q\left(\ln\left(\frac{\Phi}{p}\right)-{1\over q}\right)$, where  ${1\over p}+{1\over q}=1$; $p,q\neq 1,0$ (see e.g. RCHI, OSTI). After the field redefinition
$\phi=q\ln\left(\frac{\Phi}{p}\right)$ we arrive to $U(\phi)={1\over q}e^\phi (\phi-1)$.\medskip\\
{\bf Example 7.} Conversely, taking $U(\Phi)=e^{p W(\Phi)}\left(W(\Phi)+{1\over q}\right)$ one gets $F(R)=\left(\frac{R}{p}\right)^q\left(\ln\left(\frac{R}{p}\right)-{1\over q}\right)$.\medskip\\
{\bf Example 8.} $F(R)=R\arcsin R+\sqrt{1-R^2}$ gives $U(\Phi)=-\cos\Phi$  (cf. NI). Thus,  the inverse $\tilde{U}(\Phi)=\arccos(-\Phi) $ is obtained from $F(R)=-\sqrt{R^2-1} -\arcsin\frac{1}{R} $.\medskip\\
{\bf Example 9.} $F(R)=R\ln(R+\sqrt{1+R^2}) - \sqrt{1+R^2}$ gives $U(\Phi)=\cosh\Phi$. Thus,  the inverse $\tilde{U}(\Phi)=\arccosh\Phi $ is obtained from $F(R)=\sqrt{1+R^2} -\arcsinh\frac{1}{R} $. \medskip\\
{\bf Example 10.} In order to find out $F(R)$-lagrangian for the Higgs potential $U(\Phi)=(\Phi^2-v^2)^2$ (see DWI) one has to solve a quibic algebraic equation in the form
$$\Phi^3-v^2\Phi -\frac{R}{4}=0 .$$
Seeking real solutions of this equation one has to distinguish two cases. Either  $R>\frac{8v^3}{ 3\sqrt 3} $ and 
$$\Phi={1\over 2}\left[  R+\sqrt{R^2-{64\over 27}v^6}\right]^{1\over 3}+{1\over 2}\left[   R-\sqrt{R^2-{64\over 27}v^6}\right]^{1\over 3}=\frac{2v}{\sqrt{3}}\cosh\left({1\over 3}\ln\frac{3\sqrt 3 (R+\sqrt{R^2-{64\over 27}v^6})}{8 v^3}\right)$$
or $R<\frac{8v^3}{ 3\sqrt 3} $ and $\Phi=\frac{2v}{\sqrt{3}}\cos\left({1\over 3}\left[\arccos\frac{3\sqrt 3 R}{8 v^3}+2k\pi\right]\right)$, $k=0,1,2$.
Thus $f(R)=\int\Phi(R)\, dR$, where $\Phi$ is given by one of the 
formulas above.

Alternatively, we can start with the quadratic (Starobinsky) action and then perform
the field redefinition \eqref{e3}: $\Phi\rightarrow\Phi^2$. In such case one has to change  the other frame parameters as well (cf. \eqref{htransformations}).
\medskip\\
{\bf Example 11.} (Nojiri-Odintsov \cite{nojiri2003}) 
\footnote{In the most general form the Lagrangian is determined by $F(R)=R+ \alpha R^m+\beta R^{-n}$.}
$F(R)={1\over 2}R^2+{\alpha\over R}$. 
Then $R^3-\Phi R^{2}-\alpha=0$ and 
$$R=\frac{\Phi}{3}+\left[\frac{\Phi^3}{27}+\frac{\alpha}{2}+\sqrt{\frac{\alpha\Phi^3}{27}+\frac{\alpha^2}{4}}\right]^{1\over 3}+\left[\frac{\Phi^3}{27}+\frac{\alpha}{2}-\sqrt{\frac{\alpha\Phi^3}{27}+\frac{\alpha^2}{4}}\right]^{1\over 3}\,,$$ which in the limit $\alpha\rightarrow 0$ gives $R=\Phi$. The explict form of the potential 
$U(\Phi)= \int R(\Phi)d\Phi$ is rather complicated for $\alpha\neq 0$.\\
{\bf Example 12. } (Hu-Sawicki \cite{hu2007} ) 
\footnote{These two classes of lagrangians \cite{hu2007} and \cite{nojiri2003} are shown to pass the Solar system tests.}
This general class of  Lagrangians is given by
$F(R)=R+\frac{\alpha R^n}{\beta+R^n}\sim  -(\beta+R^n)^{-1} \leadsto U(\Phi)=\frac{\beta+(n-1)R^n}{\beta+R^n}$, where $\Phi=nR^{n-1}(\beta+R^n)^{-2}$.
Particularly, for $n=1$ we get $U(\Phi)=2\sqrt\Phi-\beta\Phi$ while for $n=2$ one has to solve the quartic equation $R^4+2\beta R^2-{2\over\Phi}R+\beta^2=0$ (see also RGI for inverse relation).\\
{\bf Example 13. } (Tsujikawa \cite{tsu2008} ) Consider $F(R)=\tanh R=\int\frac{dR}{\cosh^2 R} $ then $U(\Phi)=\int\text{arccosh}\frac{1}{\sqrt{\Phi}}\, d\Phi  = \Phi\: \text{arccosh}\frac{1}{\sqrt{\Phi}} - \sqrt{1 - \Phi}$.

\section{Transformation formulae in hybrid metric-Palatini scalar-tensor theories of gravity} \label{AppendixB}

Unlike in the metric approach, where the connection is Levi-Civita w.r.t. metric tensor, and the conformal change of the latter results in a transformation of the former, in the Palatini formalism one needs to transform these two objects separately. We postulate the following transformation formulae, defined by three functions $\gamma_i(\Phi)$ and an additional diffeomorphism of the scalar field, for the variables entering the action functional:
\begin{subequations}
	\begin{align}
		& \bar{g}_{\mu\nu}=e^{2\gamma_1(\Phi)}g_{\mu\nu}, \label{e1} \\ 
		& \bar{\Gamma}^\alpha_{\mu\nu}=\Gamma^\alpha_{\mu\nu}+2 \delta^\alpha_{(\mu}\partial_{\nu)}\gamma_2(\Phi)-g_{\mu\nu}g^{\alpha\beta}\partial_\beta\gamma_3(\Phi), \label{e2}\\
		& \bar{\Phi}=f(\Phi). \label{e3}
	\end{align}
\end{subequations}
previosly studied in \cite{kozak2019}.
When one applies these transformations to the independent variables entering the action (\ref{actionHybrid}), the functions $( {\cal A}_1,  {\cal A}_2, \mathcal{B}, \mathcal{C}_1, \mathcal{C}_2, \mathcal{V}, \alpha )$ have to be changed in the following way:
\begin{subequations}
	\begin{align}
		\mathcal{\bar{A}}_1(\bar{\Phi})&=e^{(n-2)\check{\gamma}_1(\bar{\Phi})}\mathcal{A}_1(\check{f}(\bar{\Phi})), \qquad 
		\mathcal{\bar{A}}_2(\bar{\Phi})=e^{(n-2)\check{\gamma}_1(\bar{\Phi})}\mathcal{A}_2(\check{f}(\bar{\Phi}))
		\label{ht1}\\
		\begin{split}
			\mathcal{\bar{B}}(\bar{\Phi})&=e^{(n-2)\check{\gamma}_1(\bar{\Phi})}\Big[\mathcal{B}(\check{f}(\bar{\Phi}))(\check{f}'(\bar{\Phi}))^2+(n-1)\Big(n\mathcal{A}_2(\check{f}(\bar{\Phi}))\check{\gamma}'_2(\bar{\Phi})\check{\gamma}'_3(\bar{\Phi})-\mathcal{A}_2(\check{f}(\bar{\Phi}))\left(\check{\gamma}'_2(\bar{\Phi})\right)^2\\
			&-\mathcal{A}_2(\check{f}(\bar{\Phi}))\left(\check{\gamma}'_3(\bar{\Phi})\right)^2-\frac{d\mathcal{A}_2(\check{f}(\bar{\Phi}))}{d\bar{\Phi}}(\check{\gamma}'_2(\bar{\Phi})+\check{\gamma}'_3(\bar{\Phi}))
			-2\frac{d\mathcal{A}_1(\check{f}(\bar{\Phi}))}{d\bar{\Phi}}\check{\gamma}'_1(\bar{\Phi})
			\\
			&-(n-2)\mathcal{A}_2(\check{f}(\bar{\Phi}))\check{\gamma}'_1(\bar{\Phi})(\check{\gamma}'_2(\bar{\Phi})+\check{\gamma}'_3(\bar{\Phi}))
			-(n-2)\mathcal{A}_1(\check{f}(\bar{\Phi}))(\check{\gamma}'_1(\bar{\Phi}))^2\Big)\\
			&+\check{f}'(\bar{\Phi})\Big(\mathcal{C}_1(\check{f}(\bar{\Phi}))(2n\check{\gamma}'_1(\bar{\Phi})-2(n+1)\check{\gamma}'_2(\bar{\Phi})+2\check{\gamma}'_3(\bar{\Phi}))\\
			&-\mathcal{C}_2(\check{f}(\bar{\Phi}))(2\check{\gamma}'_1(\bar{\Phi})-(n+3)\check{\gamma}'_2(\bar{\Phi})+(n+1)\check{\gamma}'_3(\bar{\Phi}))\Big)\Big],
		\end{split}
	\end{align}
	\begin{align}
		\mathcal{\bar{C}}_1(\bar{\Phi})&=e^{(n-2)\check{\gamma}_1(\bar{\Phi})}\Big[\check{f}'(\bar{\Phi})\mathcal{C}_1(\check{f}(\bar{\Phi}))-\mathcal{A}_2(\check{f}(\bar{\Phi}))\left(\frac{n-1}{2}\check{\gamma}'_2(\bar{\Phi})+\frac{n-3}{2}\check{\gamma}'_3(\bar{\Phi})\right)\Big],\\
		\mathcal{\bar{C}}_2(\bar{\Phi})&=e^{(n-2)\check{\gamma}_1(\bar{\Phi})}\Big[\check{f}'(\bar{\Phi})\mathcal{C}_2(\check{f}(\bar{\Phi}))-\mathcal{A}_2(\check{f}(\bar{\Phi}))\left((n-1)\check{\gamma}'_2(\bar{\Phi})-\check{\gamma}'_3(\bar{\Phi})\right)\Big],\\
		\mathcal{\bar{V}}(\bar{\Phi})&=e^{n\check{\gamma}_1(\bar{\Phi})}\mathcal{V}(\check{f}(\bar{\Phi})), \\
		\bar{\alpha}(\bar{\Phi})&=\alpha(\check{f}(\bar{\Phi}))+\check{\gamma}_1(\bar{\Phi})\,,\label{t6}
	\end{align} \label{htransformations}
\end{subequations}
preserving the form of the action, where $\check{f}(\bar\Phi)=\Phi$ and $\check\gamma (\bar\Phi)=-\gamma (\check f(\bar\Phi))$ 
and  $\check\gamma' (\bar\Phi)=\frac{d\,\check\gamma (\bar\Phi)}{d\,\bar\Phi} $, 
etc. (cf. \cite{kozak2019} for more detailed explanation from a group-theoretical point of view). 

In this way, the formulae above  proclaim the invariance of the action (\ref{actionHybrid}) under the transformations  \eqref{e1}-\eqref{e3} establishing the mathematical (or solution) equivalence between transformed frames. It means that changing the frame $(\ca_1, \ca_2, \cb, \cc_1, \cc_2, \cv, \alpha)$
to  $( \bar{\cal A}_1, \bar{\cal A}_2, \bar{\mathcal{B}}, \bar{\mathcal{C}}_1, \bar{\mathcal{C}}_2, \bar{\mathcal{V}}, \bar{\alpha} )$  according to \eqref{ht1}-\eqref{t6}, one should transform  solutions of the corresponding field equations by the formulas \eqref{e1}-\eqref{e3}.

One should distinguish three cases:

Setting $\mathcal{A}_2=\mathcal{C}_1=\mathcal{}C_2=0$ into the action as well as in the formulae above one reconstructs well-known metric scalar-tensor theories which contain  metric $F(R)$-subclass.
  ( the formalism introduced in \cite{kuusk2015}, slightly generalized to arbitrary dimension $n>2$ \cite{karam2017}). In this case the transformation (\ref{e2}) is not active. 

Similarly, setting $A_1=0$   one finds Palatini  scalar-tensor theories introduced in \cite{kozak2019} which contain  $F(\hr)$-subclass.

The most general case with $\ca_1, \ca_2\neq 0$ which contains hybrid $f(\hr)$-subclass has not been studied before. Moreover, it has been shown that any generalized frame $( {\cal A}_1,  {\cal A}_2, \mathcal{B}, \mathcal{C}_1, \mathcal{C}_2, \mathcal{V}, \alpha )$ is on-shell solution equivalent to the purely metric frame $\{ \bar{\cal A},  \bar{\mathcal{B}}, \mathcal{V}, \alpha \}$ with $\bar{\cal A}={\cal A}_1+{\cal A}_2$ and $\bar{\cal B}$ given by the formulae \eqref{def2}.

 Analogously to both metric and Palatini cases, it is convenient to introduce invariant quantities,
 i.e. quantities such that their functional form is independent of the conformal frame we are using:
\begin{subequations}
\begin{align}
& \mathcal{I}_\mathcal{A}(\Phi) = \frac{\mathcal{A}_1(\Phi)}{\mathcal{A}_2(\Phi)}, \\
& \mathcal{I}^{(1)}_V (\Phi)= \frac{\mathcal{V}(\Phi)}{( \mathcal{A}_1(\Phi))^{\frac{n}{n-2}}},\qquad \mathcal{I}^{(2)}_V (\Phi)= \frac{\mathcal{V}(\Phi)}{(\mathcal{A}_2(\Phi))^{\frac{n}{n-2}}}, \\
& \mathcal{I}^{(1)}_\alpha(\Phi) = \frac{\mathcal{A}_1(\Phi)}{e^{(n-2)\alpha}(\Phi)} ,\qquad
 	\mathcal{I}^{(2)}_\alpha(\Phi) = \frac{\mathcal{A}_2(\Phi)}{e^{(n-2)\alpha}(\Phi)},
\end{align}
\end{subequations}
 
 and also an integral invariant generalizing \eqref{I3} (we assume $\mathcal{A}_1(\Phi) + \mathcal{A}_2(\Phi)> 0$)
 \footnote{We have a freedom in choosing integration constant and normalization condition.} :
 \begin{equation}\label{I3g}
 \begin{split}
 &\mathcal{I}^{}(\Phi) = \int_{\Phi_0}^\Phi \frac{d\Phi'}{(\mathcal{A}_1(\Phi') + \mathcal{A}_2(\Phi'))} \Bigg| (n-1)(n-2)\mathcal{B}(\Phi') ( \mathcal{A}_1(\Phi') + \mathcal{A}_2(\Phi')) \\
 & + (1  +  \mathcal{I}^{-1}_{\ca} (\Phi'))[(n-1)\mathcal{A}'_1(\Phi')]^2   + ( \mathcal{I}_\ca (\Phi') + 1 )  [-4\mathcal{C}_1^2(\Phi') + (n^2-5)\mathcal{C}^2_2(\Phi') \\
 & 
 -2(n^2-n-4)\mathcal{C}_1(\Phi')\mathcal{C}_2(\Phi') + 2(n-1)\mathcal{A}'_2(\Phi')(\mathcal{C}_2(\Phi') - n\mathcal{C}_1(\Phi'))]
 \Bigg|^\frac{1}{2} \,.
 \end{split}
 \end{equation}
 One can notice that not all invariants are independent and some of them might be singular. However in the limiting cases ($\ca_2=0$ or $\ca_1=0$) they reproduce correspondingly the metric or Palatini ones. In fact, the integral invariant 
 can be extend further to  two parameter family ($a_1, a_2\in\mathbb{R}$):
 \begin{equation}
 \begin{split}
 &\mathcal{I}^{(a_1, a_2)}(\Phi) = \int_{\Phi_0}^\Phi \frac{d\Phi'}{(a_1\mathcal{A}_1(\Phi') + a_2 \mathcal{A}_2(\Phi'))} \Bigg| (n-1)(n-2)\mathcal{B}(\Phi') (a_1 \mathcal{A}_1(\Phi') + a_2\mathcal{A}_2(\Phi')) \\
 & + (a_1  + a_2 \mathcal{I}^{-1}_{\ca} (\Phi'))[(n-1)\mathcal{A}'_1(\Phi')]^2   + (a_1 \mathcal{I}_\ca (\Phi') + a_2  )  [-4\mathcal{C}_1^2(\Phi') + (n^2-5)\mathcal{C}^2_2(\Phi') \\
 & 
 -2(n^2-n-4)\mathcal{C}_1(\Phi')\mathcal{C}_2(\Phi') + 2(n-1)\mathcal{A}'_2(\Phi')(\mathcal{C}_2(\Phi') - n\mathcal{C}_1(\Phi'))]
 \Bigg|^\frac{1}{2} \,.
 \end{split}
 \end{equation}
%
 

\end{document}